\documentclass[aps,pra,floatfix,twocolumn,nofootinbib,superscriptaddress]{revtex4}

\usepackage{graphicx,bm,color}
\usepackage{stackengine}
\usepackage{array}
\usepackage{xcolor}
\usepackage{tabularx}
\usepackage{color}
\usepackage{amsmath}
\usepackage{amsfonts,amssymb,amsthm,stackrel,mathtools}

\usepackage{hyperref}
\hypersetup{colorlinks=true}
\hypersetup{urlcolor=blue}
\hypersetup{citecolor=black}
\hypersetup{menucolor=black}
\hypersetup{linkcolor=black}

\usepackage{cleveref}
\usepackage{longtable}
\usepackage{float}
\usepackage{pdfpages}
\usepackage[T1]{fontenc}
\usepackage[utf8]{inputenc}
\usepackage{bm}
\usepackage{ulem}
\usepackage{tikz}
\usetikzlibrary{calc,patterns,decorations.pathmorphing,decorations.markings,positioning, arrows}
\PassOptionsToPackage{hyphens}{url}
\usepackage{braket}
\usepackage{dsfont}
\usepackage{cancel}
\usepackage{soul}
\usepackage{ytableau}
\ytableausetup{textmode, boxframe=0.05em, boxsize=0.5em, centertableaux}

\newcommand{\up}{\uparrow}
\newcommand{\down}{\downarrow}

\begin{document}
\title{
Emergence of magnetic excitations in one-dimensional quantum mixtures under confinement}

\author{Pablo Capuzzi} 
\affiliation{Universidad de Buenos Aires, Facultad de Ciencias Exactas y Naturales, Departamento de Física, Pabellón 1, Ciudad Universitaria, 1428 Buenos Aires, Argentina}
\affiliation{CONICET - Universidad de Buenos Aires, Instituto de Física de Buenos Aires (IFIBA), Buenos Aires, Argentina}
\author{Patrizia Vignolo}
\affiliation{Universit\'e C\^ote d'Azur, CNRS, Institut de Physique de Nice, 06200 Nice, France}
\affiliation{Institut Universitaire de France}
\author{Anna Minguzzi}
\affiliation{Universit\'e Grenoble Alpes, CNRS, LPMMC, 38000 Grenoble, France}
\author{Silvia Musolino}
\affiliation{Universit\'e Grenoble Alpes, CNRS, LPMMC, 38000 Grenoble, France}
\affiliation{Universit\'e C\^ote d'Azur, CNRS, Institut de Physique de Nice, 06200 Nice, France}

\begin{abstract}
We obtain an exact solution for the spectral function 
for one-dimensional Bose-Bose and Fermi-Fermi mixtures with strong repulsive interactions, 
valid in arbitrary confining potentials and at all frequency scales. For the case of harmonic confinement we show that,  on top of the ladder structure of the density excitations imposed by the external confinement, spin excitations emerge as 
sideband peaks, with dispersion related to the one of 
ferromagnetic or antiferromagnetic  spin chains and a width  fundamentally larger for fermionic mixtures than for bosonic ones, as determined by the different symmetry of  spin excited states. The observation of spin excitation branches can provide a univocal probe of 
interaction-induced magnetism in ultracold atoms.
\end{abstract}
\maketitle

The understanding of the dynamics of strongly interacting quantum particles with internal degrees of freedom remains an open challenge,
due to the higher complexity of their many-body wavefunction
as compared to  the single-component case.
Quantum mixtures offer the possibility to investigate novel phases of matter, which may display pairing, magnetism, and nontrivial particle-exchange symmetries 
~\cite{Baroni2024,rev:mistakidis,rev:Sowinski2019,art:musolino2024}. 
Experiments with ultracold atomic gases
provide  an ideal platform where interactions, dimensionality, and geometry can be easily tuned~\cite{rev:Bloch,book:cohen_tannoudji_atoms}. 

Strong correlation effects and quantum fluctuations are enhanced in low dimensions. In one dimension (1D), strong repulsions forbid two particles, regardless of their spin, to occupy the same position in space, leading to fermionization. This Tonks-Girardeau (TG) regime, originally  predicted for single-component bosons~\cite{art:Girardeau1960, art:tonks1936}, 
has been generalized to mixtures of both fermions and bosons ~\cite{art:deuretzbacher2008_lett, art:volosniev_nat}. The TG  regime, experimentally achieved with ultracold atoms~\cite{art:moritz2003,art:kinoshita2004,art:paredes2004, art:kinoshita_PRL2005,art:Zurn2012}, has attracted considerable interest from a theoretical point of view~\cite{rev:cazalilla2011,rev:Guan,art:minguzzi2022strongly}, as 
its many-body wavefunction is exactly solvable, leading to exact expressions for correlation functions. Static correlations, as the one-body density matrix and the  momentum distribution ~\cite{Lenard1964,art:forrester2003,art:deuretzbacher2016_num, art:yang_2017_1bdm},
or  higher-order correlators \cite{art:Devillard2020,art:Devillard2021} 
display the emergence of universal features induced by interactions, both at short and large scales \cite{Haldane1981,art:minguzzi_PLA2002,art:olshanii_PRL2003}. 

\begin{figure}
\centering
\includegraphics[width=\linewidth]{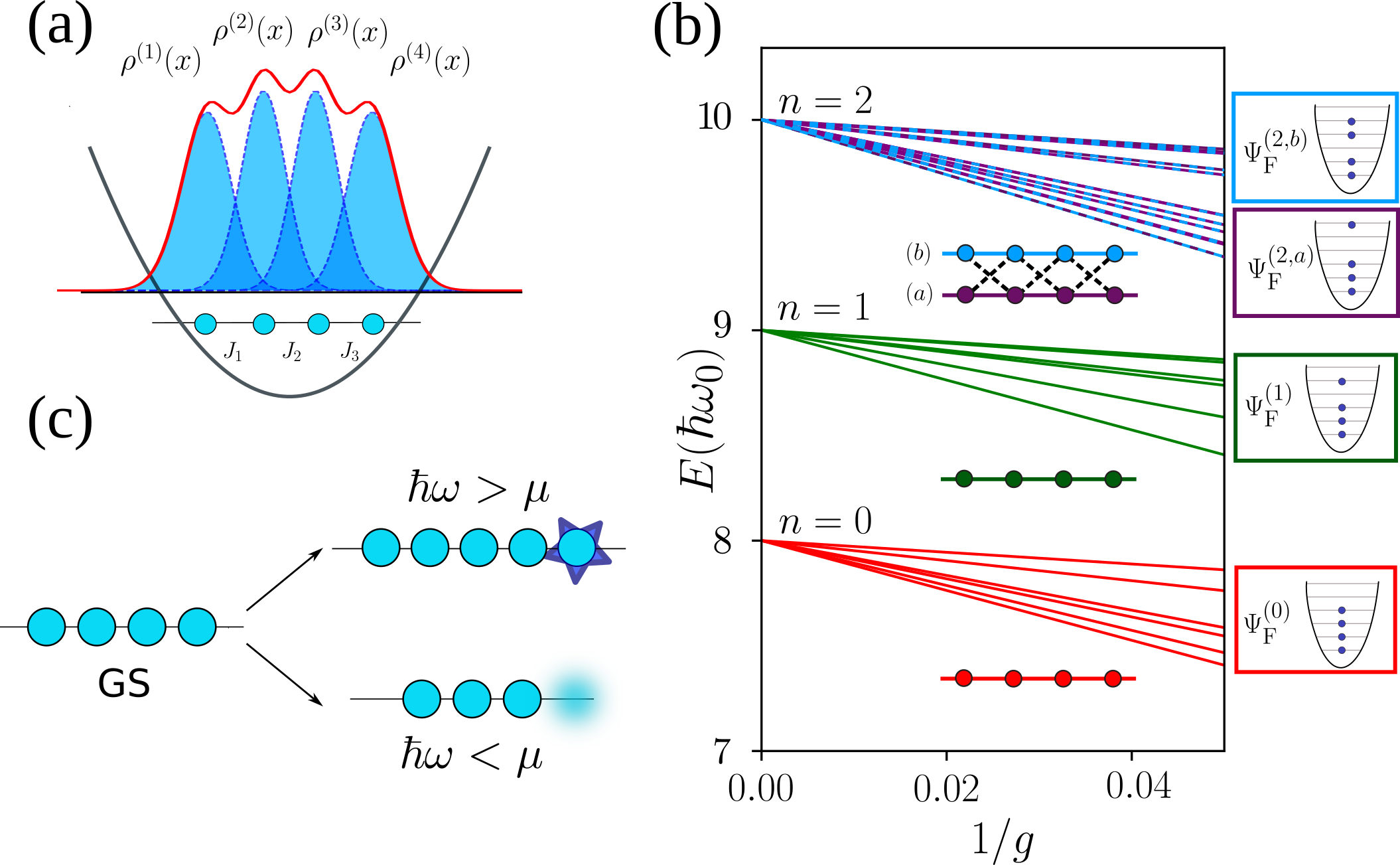}
\caption{ (a) Sketch of the excitation processes in $\mathcal{A}(k,\omega)$. (b) Sketch of the mapping of the trapped continuous system on a spin chain. (c) Ground and excited state energy manifolds for $N=4$ two-component bosons trapped in a harmonic potential as a function of the inverse interaction strength ($1/g >0$). The splitting at finite interactions is related to the spin degeneracy and the corresponding Hamiltonian can be mapped to a single or coupled spin chain, as illustrated by the sketches inside the figure. The energies at $1/g = 0$ are the Fermi energies of the corresponding non-interacting system of fermions trapped in the same external potential, whose excitations are illustrated on the right.  
}
\label{fig:model}
\end{figure}

The dynamical correlation functions contain detailed information on the excitation spectrum, which is broad even at zero temperature due to interaction effects.
We focus on
the spectral function $\mathcal{A}(k,\omega)$,  describing the response of the system upon adding or subtracting to it a particle  with momentum $\hbar k$ and energy  $\hbar \omega$ [Fig. \ref{fig:model}(a)].
The spectral function
can be measured using photoemission spectroscopy in ultracold atomic systems~\cite{art:Stewart2008,art:Vale2021,art:VincentJosse}. 
For bosons, a high-resolution spectral function has been obtained by the exact TG solution
\cite{art:settino2021,art:patu2022_finiteT, art:cheng2025}.
For a  two-component mixture at strong repulsive interactions, spin and orbital degrees of freedom decouple \cite{art:ogata_shiba,art:deuretzbacher2014,art:volosniev_nat} and the system may be viewed as  a gas of spinless hard-core particles,
with intra- and inter-component ordering fixed by an emergent Heisenberg spin chain (see Fig. \ref{fig:model}(b)). Correspondingly, its excitation spectrum is characterized by two types of low-energy excitations: charge (ie total density) and spin  excitations. 
Non-linear Luttinger liquid theory predicts the spectral structure and power-law singularities around each of those excitation branches in homogeneous systems
~\cite{art:guan2007,art:matveev2007,art:matveev2008,art:schmidt2010}. Solutions for inhomogeneous systems at strong interactions were restricted so far to the spin-incoherent regime~\cite{art:PatuFoerster2024}.  

In this Letter, we derive an exact solution for the spectral function for confined quantum mixtures at all frequency scales, allowing us to access the low-energy, spin-coherent regime of experimental relevance, containing a wealth of information on the symmetries of the spin excitations.
Starting from this theoretical foundation, 
we present, for the first time, 
how the two types of excitations coexist in  strongly repulsive 1D Bose-Bose (BB) and Fermi-Fermi (FF) mixtures 
trapped in a harmonic potential. 
We show how the excitation branches are dictated by the allowed symmetries of the initial and final states and highlight
remarkable differences among the BB and FF mixtures.

\textit{Model.} -- We consider a 1D
SU$(2)$ 
equally-balanced two-component  mixture of  $N$ particles with mass $m$, pseudo-spin $\sigma=\uparrow,\downarrow$, described by the field operator $\hat \Psi_\sigma (x)$,
confined by the external potential $V(x)$ and interacting via repulsive contact interactions of strength $g$ at zero temperature.  The Hamiltonian is given by
\begin{equation}
\begin{split}
&\hat H=  \sum_{\sigma= \up, \down} \int dx  \hat \Psi^\dagger_\sigma (x)  \left[-\frac{\hbar^2}{2m}\frac{\partial^2}{\partial x^2} + V(x)\right] \hat \Psi_\sigma (x) \\&+\frac{g}{2}  \sum_{\sigma, \sigma'} \int dx dy   \hat\Psi^\dagger_\sigma (x) \hat\Psi^\dagger_{\sigma'} (y)  \delta(x-y) \hat\Psi_{\sigma'} (y) \hat\Psi_\sigma (x),
\end{split}
 \label{eq:ham}
\end{equation}
where $V(x)= m \omega_0^2 x^2/2$ is an external  harmonic confinement with characteristic length $a_\mathrm{ho} = \sqrt{\hbar/(m \omega_0)}$. 
In the strongly repulsive limit $g\to \infty$, the many-body wave function $\Psi_{\mathbf{n}}(x_1,\sigma_1,...x_N,\sigma_N)$ 
of the multicomponent mixture
can be constructed exactly using a generalized TG Ansatz 
decoupling orbital and spin degrees of freedom
~\cite{art:deuretzbacher2008_lett, art:volosniev_nat} as
\begin{eqnarray}
\Psi_{\mathbf{n}}(x_1,\sigma_1,...x_N,\sigma_N)
= \frac{1}{\sqrt{N!}}\!\!\sum_{P\in S_N} (\pm)^P  \nonumber \\ \hat P\left(\varphi^{\mathbf{(n)}} (x_1,...x_N)  \langle {\rm Id}|\chi^{\mathbf{(n)}}\rangle \right),
\label{eq:Psi_genericn}
\end{eqnarray}
where the sum runs over the permutations $P \in S_N$
symmetry group of $N$ elements,
the orbital wavefunction $\varphi^{\mathbf{(n)}} (x_1,...x_N)\!=\!
\sqrt{N!} \theta_P(x_1, \dots, x_N) \Psi_F^{\mathbf{(n)}}(x_1, \dots, x_N)$,
with~$\theta_P(x_1, \dots, x_N)\!=\!1$ if $x_{P(1)} < x_{P(2)} < \dots < x_{P(N)}$ and zero otherwise, $\Psi_F^{(\mathbf{n})}\!(x_1, \dots, x_N)\!=\!\mathrm{det}[\phi_{n_j}(x_i)]/\sqrt{N!}$ is the Slater determinant with
energy 
$E_{\mathrm{F}}^{(n)}(N)=\hbar \omega_0 \sum_{j=1}^N (n_j+ 1/2)\!\!=\!\!\hbar \omega_0 ({N^2}/{2}+n)$,
and composed by the single-particle orbitals $\phi_{n_j}(x)$, 
solution of the one-body Schroedinger equation in
the potential
$V(x)$.
$\ket{\chi^{\mathbf{(n)}}}$ is the
spin state, described by an inhomogeneous spin-chain Hamiltonian,  as illustrated in Fig.~\ref{fig:model}(b) and (c)
\begin{equation}
\label{eq:single-chain}
\hat H_\mathrm{s} = E_{\mathrm{F}}^{(n)} \hat 1 \pm  \sum_{i=1}^{N-1}  2 J_i^{(\mathbf{n})} \!\left(  \vec{S}_{i}\!\! \cdot \vec{S}_{i+1} \mp \frac{\beta}{4}\right),
\end{equation}
where  $\vec{S}_i = (S_i^{(x)}, S_i^{(y)}, S_i^{(z)})$ are the spin operators and the top signs and $\beta = 1$ (bottom signs and $\beta = 3$) are for fermions (bosons).
Notice that spin inhomogeneity is crucial for the energy scales we are considering, while it was not taken into account in Ref. \cite{art:PatuFoerster2024} which focused on much higher scales.
Excited free-fermion many-body states  may be $p$-fold degenerate in energy, as, e.g, the second-excited state in
Fig.~\ref{fig:model}(c). In this case we use 
 Hamiltonian of $p$ coupled spin chains~\cite{art:chen2025}
wih Hilbert space dimension $M = p N!/(N_\up! N_\down!)$
\begin{equation}
\hat H_\mathrm{s}^\mathrm{(cc)} \!=\! E_\mathrm{F}^{(n)} \hat 1 \pm  \sum_{i=1}^{N-1} \sum_{\langle a, b \rangle,a,b=1}^p \!\! \!\!  \!\! 2 J_i^{(\mathbf{n}_a, \mathbf{n}_b)} \!\left(  \vec{S}_{i}^{(a)}\!\! \cdot \vec{S}_{i+1}^{(b)} \mp \frac{\beta}{4}\right),
\label{eq:coupled_spinchain}
\end{equation}
 to obtain $|\chi^{\mathbf{(n_1)}}... \chi^{\mathbf{(n_p)}}\rangle$, where $\mathbf{n}_1$...$\mathbf{n}_p$ are the quantum numbers of the degenerate many-body fermionic states and  $\langle a, b \rangle$ indicate that  $a$ and $b$ are neighbouring chains.
The hopping coefficients $J_i^{(\mathbf{n}_a, \mathbf{n}_b)}$
are 
\begin{equation}
\begin{split}
J_i^{(\mathbf{n}_a, \mathbf{n}_b)} &= \frac{N! \hbar^4}{m^2 g} \int_{x_1 < \dots < x_N} dx_1 \dots dx_N  \delta(x_i - x_{i+1})\\
&\frac{\partial [\Psi_F^{(\mathbf{n}_a)}]^\ast}{\partial x_i} \frac{\partial \Psi_F^{(\mathbf{n}_b)}}{\partial x_i}. \end{split}
\label{eq:alphai_cc}
\end{equation}
The coefficients  $J_i^{(\mathbf{n})}$  of Eq.(\ref{eq:single-chain}) are obtained by taking $\mathbf{n}_a=\mathbf{n}_b$ in Eq.(\ref{eq:alphai_cc}). 
For each free-fermion excited state, the diagonalization of the spin-chain Hamiltonian  provides the spin eigenvalues 
and eigenvectors 
needed for the calculation of the spectral function.

 \textit{Spectral function for multicomponent systems.} --
The spectral function of a many-body system contains information about the accessible energy states and their distribution in momentum space and is defined as the imaginary part of the Fourier transform of the retarded Green function~\cite{book:mahan}, i.e., for the $\sigma$ component,
\begin{equation}
\mathcal{A}_\sigma(k, \omega) = - \frac{1}{\pi} \mathrm{Im}[G^\mathrm{R}_\sigma(k, \omega)],
\label{eq:spectral}
\end{equation}
where  $G_\sigma^\mathrm{R}(k, \omega) = \int dt e^{i\omega t} \int dx dy e^{-i k(x-y)} G_\sigma^\mathrm{R}(x, t, y, t')$ is the Fourier transform of the retarded Green function, $G^\mathrm{R}_\sigma(x, t, y, t')  = \theta(t-t') (G^{>}_\sigma(x, t, y, t') - G^{<}_\sigma(x, t, y, t')) $ in terms of the lesser ($G^{<}_\sigma(x, t, y, t') = -i (\mp)\braket{\mathbf{0}| \hat{\Psi}^\dagger_\sigma(y, t') \hat{\Psi}_\sigma(x, t)| \mathbf{0}}$) and greater ($G^{>}_\sigma(x, t, y, t') = -i \braket{\mathbf{0}|\hat{\Psi}_\sigma(x,t) \hat{\Psi}^\dagger_\sigma(y,t')| {\mathbf{0}}}$) ones, with $-$ ($+$) for fermions (bosons), and
$\hat \Psi_\sigma (x, t) = e^{i \hat H t}\hat \Psi_\sigma (x) e^{-i \hat H t}$.  As  illustrated schematically in Fig.~\ref{fig:model}(b),
the lesser (greater) Green's function contains information about the process of  removing (adding)  a particle to the system and contributes for energies $\hbar \omega < \mu$ ($\hbar \omega > \mu$) with $\mu$ the chemical potential.

Following the notation of~\cite{book:korepin1993, art:patu_SF2comp}, we express the lesser and greater Green's functions as a sum over form factors,  $F_\sigma^{(\mathbf{n}_1, \mathbf{n}_2)} (x, t) = \braket{\mathbf{n}_1 | e^{i\hat H t} \hat \Psi_\sigma(x) e^{-i\hat H t} | \mathbf n_2}$ with ${\bf n}_1$ and ${\bf n}_2 $ sets of quantum numbers, namely,
 \begin{eqnarray}
 \mp i G^{<}_\sigma (x, t , y, t') &=& \sum_{\mathbf{n}} [F_\sigma^{(\mathbf{n}, \mathbf{0})} (y, t')]^\ast F_\sigma^{(\mathbf{n}, \mathbf{0})} (x, t),  \label{eq:glesser_ff} \\
 i G^{>}_\sigma (x, t , y, t') &=& \sum_{\bf m} F_\sigma^{({\bf 0 }, {\bf m})} (x, t) [F_\sigma^{({\bf 0 }, {\bf m})} (y, t')]^\ast ,
\label{eq:ggreater_ff}
\end{eqnarray}
where $\mathbf{n}$ and $\mathbf{m}$ collect the indices for the complete set of the excited states with $N-1$ and $N+1$ particles, respectively and $\mathbf{0}$ indicates the ground state.
Using 
Eq.~\eqref{eq:Psi_genericn},  similarly to the one-body density matrix~\cite{art:deuretzbacher2016_num}, the form factors in Eqs.~\eqref{eq:glesser_ff} and~\eqref{eq:ggreater_ff} can be factorized into a spatial and spin part, as follows,
\begin{eqnarray}
 F_\sigma^{({\bf n}, \mathbf{0})} (x, t) &=& \sum_{j=1}^N  (\pm 1)^{1+j}  g_{<}^{(1, j)} (\mathbf{n} | x, t) \eta^{(1, j)}_< (\mathbf{n} | \sigma, t),\,\,\,\,\,\,\,\,\,\,
\label{eq:Fdef_Pij}\\
 F_\sigma^{(\mathbf{0}, {\bf m})} (x, t) &=& \sum_{j=1}^{N+1}  (\pm 1)^{1+j}  g_>^{(1, j)} (\mathbf{m} | x, t) \eta_>^{(1, j)} (\mathbf{m} | \sigma, t),\,\,\,\,\,\,\,\,\,\,
\label{eq:Fdef_Pij_bar}
\end{eqnarray}
where both spatial and spin  parts quantifies the overlap between the ground state of $N$ particles and the excited state of $N-1$ particles, for the lesser component, and $N+1$ particles for the greater component.    The expressions for 
$ g_{\lessgtr}^{(1, j)} (\mathbf{m} | x, t)$ and $\eta_{\lessgtr}^{(1, j)}$ are  given in~\cite{suppmat}.
Equations (\ref{eq:glesser_ff})-(\ref{eq:Fdef_Pij_bar}) generalize to the confined and continuous case the result obtained on a uniform lattice 
in ~\cite{Parola1992,Penc1997}.

\begin{figure*}
\includegraphics[width=0.95\textwidth]{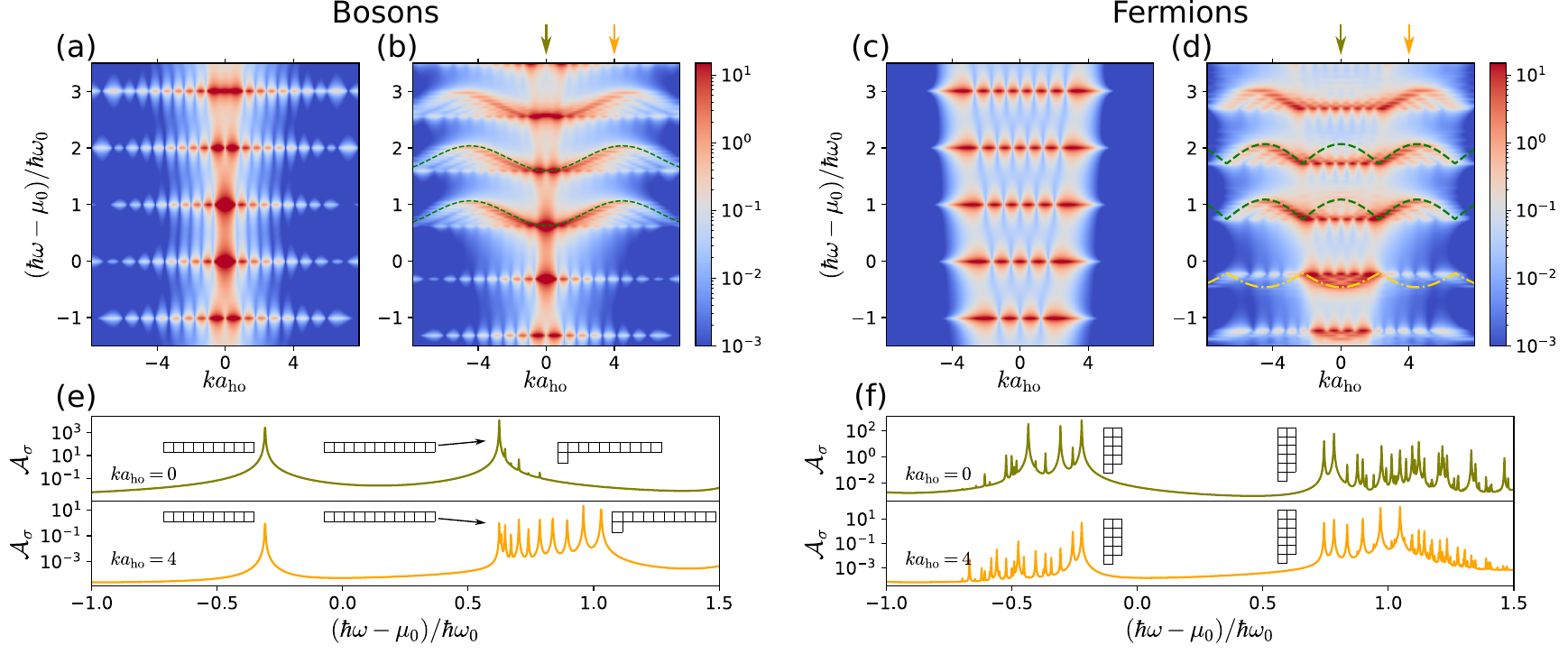}
\caption{Top panels: spectral function of bosonic and fermionic gases in harmonic potential, in the $k-\omega$ plane (a) $N=10$ Tonks-Girardeau bosons (b) $N=5+5$ strongly repulsive BB mixture $(g = 100\hbar\omega_0\,a_{\mathrm{ho}})$ .
(c) $N=5$ spinless, noninteracting  fermions 
(d)  $N=5+5$ strongly repulsive FF mixture   $(g = 100\hbar\omega_0\,a_{\mathrm{ho}})$. The dashed lines in panels (b) and (d) are the theoretical prediction  for ferromagnetic (antiferromagnetic) spin chains for bosonic (fermionic) mixtures respectively.  Bottom panels:
 cuts of $\mathcal{A}_\sigma(k,\omega)$ at fixed $k$ values ($k a_\mathrm{ho} = 0$ top and  $k a_\mathrm{ho} = 4$ bottom), indicated by the arrows in (b) and (d).  The Young diagrams indicate the symmetry of the excited states.   
The zero of the frequency is set to 
 $\mu_0 = E_{\mathrm{F}}^{(0)}(N) - E_{\mathrm{F}}^{(0)}(N-1) = \hbar \omega_0(N\!-\!\frac{1}{2})$,  chemical potential of the mapped Fermi gas.
}
\label{fig:spectral_BF}
\end{figure*}
  
\textit{Spectra for BB and FF mixtures in harmonic trap.} --
Here, we illustrate our method by calculating the exact spectral function for
$N= 10$ two-component bosons and fermions trapped in a harmonic potential. 
 We analyze first the spectral functions for single component systems, either
 bosons  in the TG limit $g\to \infty$ 
 or 
 non-interacting fermions, 
 shown on panels (a) and (c) of  Fig.~\ref{fig:spectral_BF},
 respectively.  In both cases, spectral peaks are located at energies multiples of $\hbar \omega_0$. 
For non-interacting fermions, this readily follows from the
 analytical expression of the  spectral function  $\mathcal{A}(k, \omega) = \sum_{n_i} \delta(\omega - \epsilon_{n_i}) |p_{n_i}(k)|^2$ with  $p_{n} (k) = \int dx e^{ik x} \phi_n (x)$ 
the Fourier transform of the single-particle orbital, showing that it has spectral peaks at the harmonic-oscillator energies $\epsilon_{n_i}=\hbar \omega_0 (n_i+1/2)$. The momentum weights   $|p_{n_i}(k)|^2$ determine the size and multiplicity of the peaks (see also 
~\cite{art:yamamoto2011} for  fermions  in optical lattices). Also for TG bosons, excitation energies are the same as the one of free fermions. These correspond to  phonon (also called holons \cite{art:matveev2007})  modes, discretized due to the harmonic confinement \cite{art:hydro2001}, thus highlighting the collective nature of excitations in 1D interacting systems. As compared to free fermions, TG bosons have a much smaller momentum spread, similarly to what happens in the momentum distribution, to which the spectral function is related.

The discretized phonon modes, with energy multiples of $\hbar \omega_0$, are also visible in the spectral function  for strongly interacting multicomponent mixtures, shown in  panels (b) and (d) of Fig.~\ref{fig:spectral_BF}.  On top of such charge excitations, these spectra clearly show  the emergence of additional  dispersive branches,  which we associate to spin excitations (spinons): this is confirmed both by the overall shape of the dispersion and by the analysis of the number of spectral peaks. 
We first notice that the bending of the spectral function in Fig.~\ref{fig:spectral_BF}(b) and (d) follows the energy dispersion of the low-energy excitations of a uniform ferromagnetic or antiferromagnetic spin chain, respectively, indicated by the dashed lines.

Specifically, for BB mixtures
the observed spin dispersion in Fig.~\ref{fig:spectral_BF}(b) agrees well with 
\begin{equation}
   \!\!\! \hbar \omega_{\mathrm{FM}}(k) =  \epsilon_\mathrm{FM}^{(0)}(N+1)- \epsilon_\mathrm{FM}^{(0)}(N) +2 J_c^{(\bf n)}\left[\cos (k/n_\mathrm{max}) - 1\right] 
    \label{eq:line>_B}
\end{equation}
 where $\epsilon_\mathrm{FM}^{(0)}(N) = - 2 \sum_{i=1}^{N-1} J_i^{(\mathbf{n})}$ is the ground-state energy of the ferromagnetic spin Hamiltonian~\eqref{eq:single-chain}
 associated to the corresponding energy manifold
 and the last term is the dispersion of the lowest-energy excitations of a ferromagnet,
  which are spin waves induced by flipping a single spin
  with respect to the ground state~\cite{book:parkinson_spinsystems} and the
  hopping $J_c^{(\bf n)}$ is the one evaluated at the center of the trap.
  Notice that since the spins are attached to the particles that move in real space 
 ~\cite{art:matveev2008}, the physical 
 momentum $k$ is related to the wave number $Q$ of the
 spin-chain excitation according 
 to $k=Q n_\mathrm{max}$, with  $n_\mathrm{max} = k_F/\pi$  the density at the center of the trap. 
 The above expression (\ref{eq:line>_B}) well describes the two lowest-energy spin excitation branches of bosons of Fig. \ref{fig:spectral_BF} occurring  for $\hbar \omega > \mu$, while for the higher branches the dispersion is not available analytically since it involves coupled spin chains.
 
For FF mixtures, the observed spin dispersions in  Fig.~\ref{fig:spectral_BF}(d) are in excellent agreement with  the two spin excitation branches $\omega^\mp_{\mathrm{AFM}}(k)$ expected for energies above and below $\mu$ respectively,
\begin{equation}
   \hbar\omega^\mp_{\mathrm{AFM}}(k)\!\! =\!\! \pm\!\left[\!\epsilon_\mathrm{AFM}^{(0)}\!(N)\!-\!\epsilon_\mathrm{AFM}^{(0)}\!(\!N\!\mp\!1\!)\!-\!\pi  J_c^{(\bf n)}\!\!\sin  \left|\frac{k}{n_\mathrm{max}} \!-\!\frac{\pi}{2}\right| \right]
    \label{eq:line>_F}
\end{equation}
where  $\epsilon_\mathrm{AFM}^{(0)}(N) = - 2\mathrm{ln} 2\sum_{i=1}^{N-1} J_i^{(\mathbf{n})}$ is the ground-state energy of the antiferromagnetic spin chain ~\eqref{eq:single-chain} ~\cite{art:bethe1931, art:hulthen1938} associated to the corresponding energy manifold, and the last term is the des Cloizeaux and Pearson  excitation energy of an antiferromagnet~\cite{art:desCloizeaux1962},  with hopping $J^{(\bf n)}_c$. We notice that for fermions the wave number is shifted by $\pi/2$, because the fermionic excitations occur around the Fermi momentum of each spin component  $k_{F,\sigma}=k_F/2$, 
~\cite{Penc1997, art:Yamanaka1997, Soltanien2014}. Finally,  we observe that the lesser and greater contributions to  the spectral function are nonzero in alternating momentum regions,  consistently with the predictions of the non-linear Luttinger liquid theory~\cite{art:imambekov_rev}.

Remarkably, both for BB and for FF mixtures the agreement of the calculated spin spectral lines with   Eqs.~(\ref{eq:line>_B})-(\ref{eq:line>_F}) is excellent even in the presence of the harmonic trap:
this is due to the fact that the largest contribution to the spectral function is given by the high-density region at the center of the trap, where  the density varies only slightly.

We proceed next to demonstrate that the multiplicity of the spectral peaks is fully understood by analyzing the allowed spin excitations. The spin Hamiltonian (\ref{eq:coupled_spinchain}) is   SU(2)-symmetric and its eigenstates have definite symmetry under particle exchange,
as characterized by Young diagrams, which we denote as $(a,b,c,...)$ indicating the number of boxes in each line of the diagram. By looking at the symmetry of the initial state and the one obtained upon adding or subtracting one particle to the system, we explain the number of peaks observed in the spectral function  for a given spin band shown in  Fig.~\ref{fig:spectral_BF}(e)-(f).

For the BB mixture, the ferromagnetic  ground state
has  total spin quantum number
$S=N/2$~\cite{art:eisenberg2002, art:yang2003}, and
$S_z= 0$ in  the equally balanced case we are considering. The Young diagram corresponding to this state is the fully symmetric one,  given by  a single line with $N$ boxes  ($N$).
 For $\hbar \omega < \mu$, the lower-energy branches correspond to excited states with ($N-1$) symmetry, 
  $S=(N -1)/2$ and $S_z= \pm 1/2$, which are non-degenerate, yielding a single peak in Fig.~\ref{fig:spectral_BF}(e) for $\hbar \omega < \mu$.  Instead, for $\hbar \omega >\mu$, we have two possible symmetries of the excited state: the line diagram ($N+1$),  that
has dimension 1
 and the angle diagram ($N,1$) that has dimension $N$
 with $S=(N+1)/2$ and $S=(N-1)/2$ respectively. 
 This multiplicity emerges in the spectral cuts of 
 Fig.~\ref{fig:spectral_BF}(e)  for  $\hbar\omega>\mu$, where we observe  in total $N+1$ peaks in a given spin excitation band, with the lowest-energy one given by the line diagram. 
 
For the FF mixture the antiferromagnetic ground state has   $S = 0$ and $S_z = 0$~\cite{art:LiebMattis1962} and $(2,2,..., 2)$ symmetry.. 
This means that, during the process of particle removal (particle addition), we can only obtain the  $(2,2,...1)$ diagram  (($2,2,...2,1$) diagram) with dimension $D_-$ ($D_+$), where
$D_{\pm}=2(N\pm 1)!/(((N-1\pm1)/2)!((N+3\pm1)/2)!)$ 
corresponding in both cases to 
$S = 1/2$ and $S_z = \pm 1/2$. This leads, for $N=10$, to $42$ peaks for the first branch at $\hbar\omega<\mu$
and to $132$ peaks for the one at $\hbar\omega>\mu$, as shown
in Fig.~\ref{fig:spectral_BF}(f).
For higher branches stemming from  degenerate excited-state manifolds, the number of peaks of the spin band are multiplied by $p$. 

 \begin{figure}
    \centering
    \includegraphics[width=0.85\linewidth]{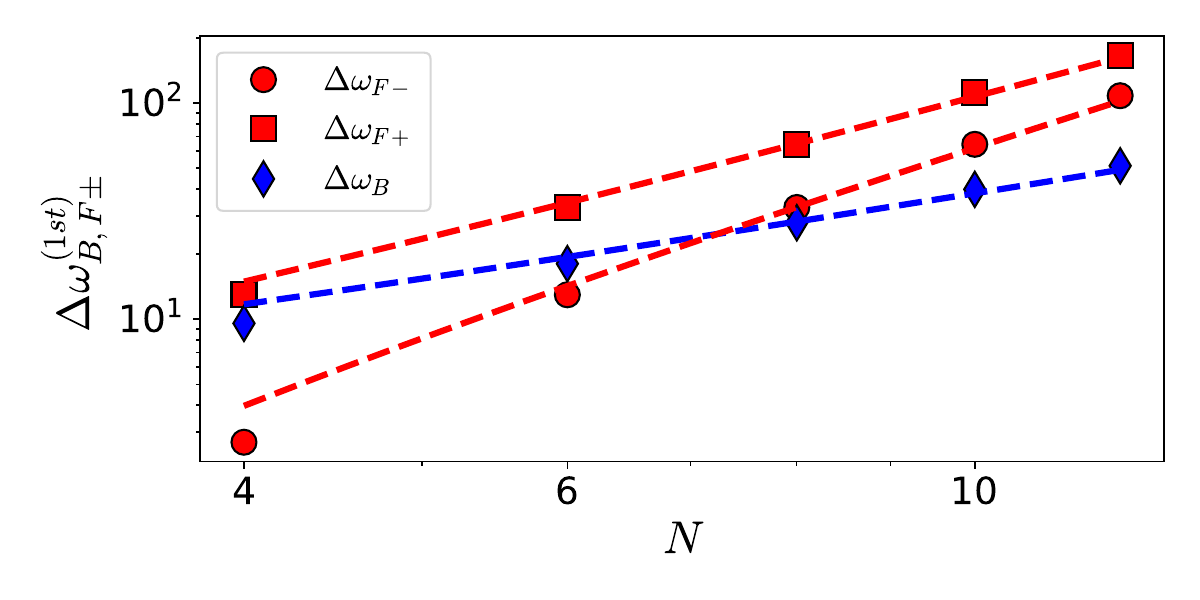}
    \caption{Energy widths $\Delta\omega_{B,F\pm}^{(1st)}$  of the lowest spin excitation bands  of bosonic (B) and fermionic (F)  mixtures in a harmonic trap,  where in the latter case  $+(-)$ refer to excitations with energies above (below) $\mu$, respectively,
    as functions of particle number $N$, from the exact spectral function (symbols) and large-$N$ analytical predictions (dashed lines).
    \label{fig:spre}}
\end{figure}

Finally, we estimate the width of the bosonic and fermionic   $n$-th  spin excitation bands for arbitrary $N$ by making an Ansatz on the corresponding spin excited states \cite{art:decamp2016_high,AupetitDiallo2024,suppmat}. For the BB mixture, spin excitations occur only  at $\hbar\omega>\mu$ and both highest and lowest-energy excitations are associated to highly symmetric states (see again Fig.~\ref{fig:spectral_BF}), yielding   $\Delta\omega^{(\mathbf{n})}_{\rm B} \sim 4 {J}^{({\bf n})}_{c}/\hbar$. For
  the FF mixture all spin excitations are associated to a more antisymmetric diagram and their highest- and lowest excited states scale differently with $N$, leading to  $\Delta\omega^{({\bf n})}_{\rm F}\!\! \sim (N-2\pm 1){J}^{({\bf n})}_{c}/\hbar$   with $+(-)$  referring to energies larger (smaller) than $\mu$. 
Using that the coefficient  $J \propto n^3$  ~\cite{art:deuretzbacher2014}, and that at the center of the trap $n_{\rm max}\simeq\sqrt{2N}/(\pi a_{\rm ho})$ for the first excited branches, we predict at $N\gg 1$  that $\Delta\omega^{(1st)}_{\rm B}\propto (N+1)^{3/2}$,
$\Delta\omega^{(1st)}_{ \rm F\pm }\propto (N\pm1)^{5/2}$.
 In Fig.~\ref{fig:spre} we show that our estimate works very well for the spectra obtained by the exact solution, already for small particle numbers.
Finally, we remark that the mean distance between two consecutive peaks $\delta\omega^{({\bf n })}=\Delta\omega^{({\bf n})}/N_{\rm peaks}^{({\bf n})}$ 
scales as $\delta\omega^{(1st)}_{\rm B}\propto\sqrt{N}$ and $\delta\omega^{(1st)}_{\rm F\pm}\propto N^4 2^{-(N+1)}$ for bosons and fermions at large $N$, hence it will be experimentally easier to resolve the individual spin excitations
with a BB mixture rather than with a FF one.

\textit{Conclusions.} --  By exploiting separation of spin and orbital degrees of freedom at strong coupling,  we have obtained an exact expression for 
the spectral function of strongly repulsive binary quantum mixtures confined by arbitrary external potential allowing for high-precision many-body calculations.
 In the case  Bose-Bose and Fermi-Fermi equally-balanced 
 mixtures under harmonic confinement,  we find that spin excitations clearly emerge in the spectral function on top of each quantized phonon mode
 determined by the harmonic potential. This provides a clear indication on how  to access and probe spin excitations in experiments with ultracold atoms.
 Moreover, we have demonstrated that each spin excitation branch has a dispersion that closely follows the one of uniform ferromagnetic and antiferromagnetic  spin chains, thereby yielding a univocal signature of the emergence interaction-induced magnetism in such systems. Finally we 
  highlight that a
 peculiarity of the harmonic confinement is that, at high energy,   excitations with different orbital quantum numbers may contribute to the same excitation branch. This effect, that we have described using a mapping to
 coupled spin chains, paves the way to the
 study of spin excitations in quasi-1D systems.

\begin{acknowledgments}
We thank F. H\'ebert,  M. Albert and W.-J. Chetcuti  for valuable discussions. We acknowledge financial support from the ANR-21-CE47-0009 Quantum-SOPHA project, 
from the ANR-23-PETQ-0001 Dyn1D at the title of
France 2030 and  from the CNRS International Research Project COQSYS.
\end{acknowledgments}

\bibliographystyle{apsrev4-2}
\bibliography{biblio-arxiv}

\appendix

\begin{widetext}

\section{Detailed expressions of form factors}
We provide here the full expressions for the orbital and spin part of the form factors, i.e.
$g_{<}^{(1, j)}$, $g_{>}^{(1, j)}$,  and $\eta^{(1, j)}_<$ $\eta^{(1, j)}_>$ entering Eqs.~(9) and (10) of the main text. They directly follow from the definition of lesser and greater Green's functions, Eqs.~(7) and (8) of the main text. 
The orbital parts of the form factors are given by
\begin{align}
 &g_{<}^{(1, j)}\!(\mathbf{n} | x, t)\! =\!  (N\!\!-\!\!1)!\!\!\! \int_{x_2 < \cdots x_{j} < x < x_{j+1} < \cdots < x_N}\!\!\!\!\!\!\!\!\!\!\!\!\!\!\!\!\!\!\!\!\!\!\!\!\!\!\!\!\!\!\!\!\!\!\!\!\!\!\!\!\!\!\!\!\!\!\!\!\!\!\!\!\!\!\!\!\! d\vec{X}_{N\!-\!1}   [\Psi_F^{(\mathbf{n})}\!(\vec{X}_{N\!-\!1}, t)]^\ast \Psi_F^{(0)}\!(x, \vec{X}_{N\!-\!1}, t),
\label{eq:gn_1j} \\
&g_>^{(1, j)}\!(\mathbf{m} | x, t)\!=\! N! \int_{x_2 < \cdots x_{j} < x < x_{j+1} < \cdots < x_{N+1}}\!\!\!\!\!\!\!\!\!\!\!\!\!\!\!\!\!\!\!\!\!\!\!\!\!\!\!\!\!\!\!\!\!\!\!\!\!\!\!\!\!\!\!\!\!\!\!\!\!\!\!\!\!\!\!\!\! d\vec{X}_{N} [\Psi_F^{(\mathbf{m})} (x, \vec{X}_{N}, t)]^\ast \Psi_F^{(0)} (\vec{X}_{N}, t),
\label{eq:gn_1j_bar}
\end{align}
with  $\vec{X}_{N} = \{x_i \, |\, x_i \neq x\}$ sets of $N$ position coordinates,
and the spin parts by
\begin{align}
\eta^{(1, j)}_< (\mathbf{n} | \sigma, t)  &=\!\!\! \sum_{\vec{\sigma}_{N-1}}\!\! \braket{\chi^{(\mathbf{n} )}(t) | \vec{\sigma}_{N-1}}   \braket{\sigma, \vec{\sigma}_{N-1} | \hat P_{(1, \dots, j)} | \chi^{({\bf 0})}(t)}\nonumber\\
&= \!\!\!\!\!\!\sum_{P \in \Sigma_{N-1}}\!\!\![a_P^{(\mathbf{n} )} (t)]^\ast a_{P_{(1 \dots, j)}}^{({\bf 0})} (t)\delta_{\sigma, \sigma_1},   
\label{eq:eta_1j} \\
\eta^{(1, j)}_> (\mathbf{m} | \sigma, t)  &= \sum_{\vec{\sigma}_{N}} \!\! \braket{\chi^{(\mathbf{m} )}(t) | \hat P_{(j, \dots, 1)} |\sigma, \vec{\sigma}_{N} }   \braket{ \vec{\sigma}_{N} | \chi^{({\bf 0})}(t)}\nonumber\\
&= \!\!\!\!\!\!\sum_{P\in \Sigma_N+1}\!\!\![a_{P_{(j \dots, 1)}}^{({\bf m})} (t)]^\ast a_{P}^{({\bf 0})} (t) \delta_{\sigma, \sigma_1},   
\label{eq:eta_1j_bar}
\end{align}
with $\vec{\sigma}_{N} = \{\sigma_i \, |\, \sigma_i \neq \sigma\}$  sets of $N$ spins, which are attached to the position coordinates $\{x_1,...x_N \}$.  The pair of indices $(1, j)$ in Eqs.~\eqref{eq:gn_1j}-\eqref{eq:eta_1j_bar} represents a fixed ordering of the particles, as indicated by the integration ranges in the spatial parts and by the cycle (anti-cycle) $P_{(1, \dots, j)}$ ($P_{(j, \dots, 1)} = P^\dagger_{(1, \dots, j)}$), i.e., $1 \to 2 \to \dots \to j-1 \to j \to 1$ ($j \to j-1 \to \dots \to 2 \to 1 \to j$). Above, we have used that $|\chi^{(\mathbf{n})}(t)\rangle= e^{-i H_\mathrm{s} t}|\chi^{(\mathbf{n})}(0)\rangle$ and $\Psi_F(\vec X, t)=\det[\phi_j(x_\ell,t)]$ with the time evolved single-particle orbitals $\phi_j(x_\ell,t)=e^{-i \epsilon_j t}\phi_j(x_\ell,0)$. In the case of degenerate free-fermion excited states, we use instead $|\chi^{(\mathbf{n}_1)}\ldots\chi^{(\mathbf{n}_p)}(t)\rangle= e^{-i H_\mathrm{s}^{\mathrm{(cc)}} t}|\chi^{(\mathbf{n}_1)}\ldots\chi^{(\mathbf{n}_p)}(0)\rangle$ and 
the linearity of the multi-chain states allows us to
compute the values of corresponding $g_<$ and $g_>$ and $\eta_<$ and $\eta_>$ as simple summations.

\section{Lesser Green's function: Spatial part of form factors}\label{app:spatial_lesser}
In this section, we derive a compact form to calculate Eq.~\eqref{eq:gn_1j} 
by following similar steps used in Ref.~\cite{art:deuretzbacher2016_num} to derive the one-body density matrix. The first simplification comes from the fact that the integrand in Eq.~\eqref{eq:gn_1j}  is symmetric under permutations of $x_2, \dots, x_{j}$ and $x_{j+1}, \dots, x_{N}$, and  therefore we obtain
\begin{equation}
\begin{split}
g_{<}^{(1, j)} (\mathbf{n} | x, t) &= \frac{(N-1)!}{(j-1)! (N-j)!} \sum_{i=1}^N (\pm 1)^{i+1}\phi_{0_i} (x, t)\int_{x_2, \cdots, x_{j} < x < x_{j+1}, \cdots,  x_N}\!\!\!\!\!\!\!\!\!\!\!\! d\vec{X}_{N-1} [\Psi_F^{(\mathbf{n})} (\vec{X}_{N-1}, t)]^\ast  \Psi_F^{(\{0_k | 0_k\neq 0_i\})} (\vec{X}_{N-1}, t),
\end{split}
\label{eq:spatial_G<}
\end{equation}
where we have removed the row dependent on $x$ in the Slater determinant of $N$ particles which means that if $0_k$ are the quantum numbers associated to the ground state, we have to remove the one equal to $i$, as indicated in the subscript of the wave function. The Slater determinant can be rewritten using the  Leibniz representation, i.e., $\Psi_F(\vec{X}_{N-1}, t) = (1/\sqrt{(N-1)!}) \sum_Q (-1)^Q \prod_{i=2}^N \phi_{Q(i)} (x_i, t)$ and Eq.~\eqref{eq:spatial_G<} becomes
\begin{equation}
\begin{split}
g_{<}^{(1, j)} (\mathbf{n} | x, t)  &= \frac{1}{(j-1)! (N-j)!}\sum_{i=1}^N (-1)^{i+1}\phi_{0_i} (x, t) 
\int_{x_2, \cdots, x_{j} < x < x_{j+1}, \cdots,  x_N}\!\!\!\!\!\!\!\!\!\!\!\! d\vec{X}_{N-1}\sum_{Q, Q' \in S_{N-1}} (-1)^{Q+Q'}\\
&\times \left(\prod_{l=2}^N [\phi_{Q(n_l)} (x_l, t)]^\ast \right) \left(\prod_{k=2}^N \phi_{Q'(0_{k} \neq 0_{i})} (x_{k}, t) \right),
\end{split}
\label{eq:g1l}
\end{equation}
where in the last product of single-particle orbitals, for each $i$, we need to relabel the quantum numbers related to the ground state, such that $\{ 0_1, \dots, 0_{i-1}, 0_{i+1}, \dots, 0_N \} \to \{0_2, \dots, 0_{N}\}$.

Following Ref.~\cite{art:deuretzbacher2016_num}, we  introduce the 1D integrals
\begin{equation}
\begin{split}
    A_{i, j}(x, t) &= \int_{-\infty}^x dz [\phi_{i} (z, t)]^\ast \phi_{j} (z, t) \\
    &=\delta_{i, j} - B_{i, j}(x, t),
    \end{split}
    \label{eq:Aij_g}
\end{equation}
with, due to the orthogonality relation between single-particle orbitals,
\begin{equation}
\begin{split}
B_{i, j}(x, t) &= \int_{x}^\infty dz [\phi_{i} (z, t)]^\ast \phi_{j} (z, t) \\
&= e^{-i(\epsilon_j - \epsilon_i) t} \int_{x}^\infty dz [\phi_{i} (z, 0)]^\ast \phi_{j} (z, 0) 
\end{split}
\label{eq:Bij_g}
\end{equation}
where $\epsilon_i$  is the energy
level corresponding to the orbital $\phi_i(x)$.  We therefore rewrite Eq.~\eqref{eq:g1l} as follows
\begin{equation}
\begin{split}
g_{<}^{(1, j)} (\mathbf{n} | x, t)  &= \frac{1}{(j-1)! (N-j)!} \sum_{i=1}^N (-1)^{i+1}\phi_{0_i} (x, t) 
\sum_{Q, Q' \in S_{N-1}} (-1)^{Q+Q'} A_{Q(n_2), Q'(0_2)}(x, t) \cdots
A_{Q(n_{j}), Q'(0_{j})}(x, t)\\ &\times B_{Q(n_{j+1}), Q'(0_{j+1})}(x, t) \cdots B_{Q(n_{N}), Q'(0_{N})}(x, t),
\end{split}
\label{eq:g1l_AB}
\end{equation}
with $n_j \geq n_2$ and $\{0_2, \dots, 0_{N}\} \neq 0_i$. We then introduce the permutation $Q''$ such that $Q'=Q'' \circ Q$ and the Left and Right disjoint subsets of indices, $L = \{ Q(2), \dots, Q(j) \}$ and $R = \{ Q(j+1), \dots, Q(N) \}$, as sketched in Fig.~\ref{fig:subsets}. We substitute the sum over $Q$ with a sum over $L+R$ and therefore Eq.~(\ref{eq:g1l_AB}) becomes
\begin{equation}
\begin{split}
g_{<}^{(1, j)} (\mathbf{n} | x, t)  &= \sum_{i=1}^N (-1)^{i+1}\phi_{0_i} (x, t) \sum_{Q''} (-1)^{Q''} \sum_{L+R = \underline{N-1}} \prod_{l \in L} A_{n_l, 0_{Q''(l)}}(x, t) \prod_{r \in R} B_{n_r, 0_{Q''(r)}}(x, t) 
\end{split}
\label{eq:g1l_AB_1}
\end{equation}
where we have multiplied the result by $(j-1)! (N-j)!$ because the sum over $L+R$ contains  $(j-1)! (N-j)!$ and the ordering among $A$s and $B$s is not important~\cite{art:deuretzbacher2016_num}. The notation $\underline{N-1}$ indicates the ordered set $\{2, \dots, N\}$.
\begin{figure}
\centering
\includegraphics[scale=0.4]{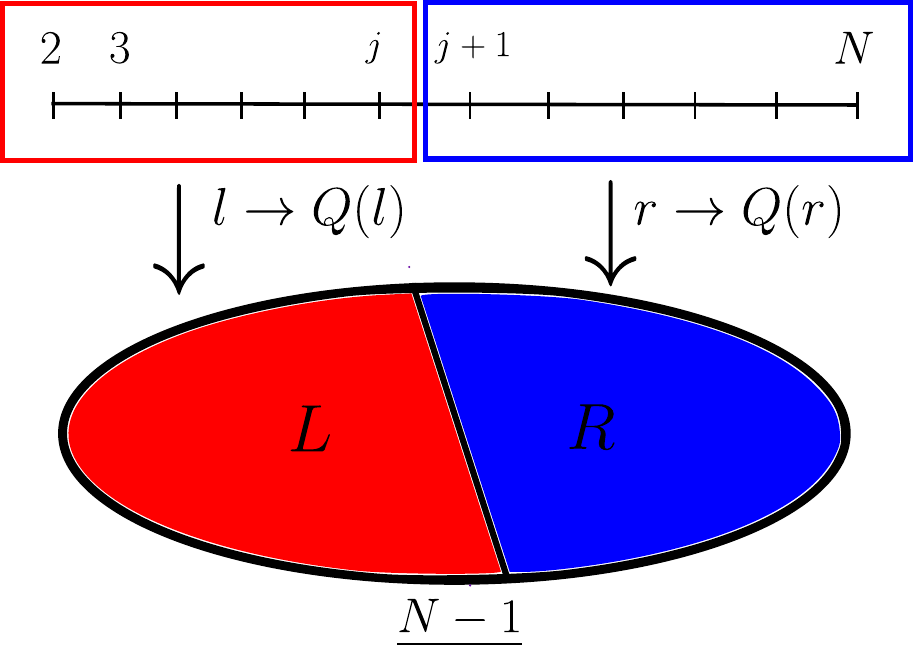}
\caption{Schematic representation of the subsets in Eq.~\eqref{eq:g1l_AB_1}. The starting ordered set of indices $\{2, \dots, N\}$  is represented by the horizontal line. When we apply the permutation $Q$, the indices $l \in [2, j]$ go to the subset $L$, instead the indices  $r \in [j+1, N]$ go to the subset $R$. The subset $L$ contains two disjoint subsets, $L_1$ and $L_2$, such that if $Q(l) \in L_1$ if $Q''(Q(l)) = Q(l)$ and all the remaining indices belong to $L_2$.}
\label{fig:subsets}
\end{figure}

The goal is now to rewrite the product of $A$s in terms of $B$s in order to obtain a determinant as final expression. We therefore rewrite the product of $A$s in terms of $B$s using the following relation~\cite{art:deuretzbacher2016_num}
\begin{equation}
\begin{split}
\prod_{l \in L}  A_{n_l, 0_{Q''(l)}}(x, t) &=  \sum_{L_1+L_2 =L} (-1)^{|L_1|+|L|} \prod_{l\in L_2} B_{n_l, 0_{Q''(l)}} (x, t)
\end{split}
\label{eq:prodAB}
\end{equation}
where $|L|$ indicates the cardinality of the subset $L$, which has been decomposed in the disjoint subsets $L_1$ and $L_2$. The elements of $L_1$ are such that are mapped onto themselves by $Q''$ (if $l \in L_1$, then $Q''(l) = l$). Importantly, the sum over $L_1+L_2 = L$ guarantees that the size of the index space remains intact for each element. For example, if $L$ has three elements, $\{Q(2), Q(3), Q(4)\}$, all the three indices will be present for each element of the product, as indices of the $B$s ($\in L_2$) or $\delta$s ($\in L_1$). 

Using Eq.~\eqref{eq:prodAB}, Eq.~\eqref{eq:g1l_AB_1} becomes
\begin{equation}
\begin{split}
g_{<}^{(1, j)} (\mathbf{n} | x, t)  &= \sum_{i=1}^N (-1)^{i+1}\phi_{0_i} (x, t)\sum_{Q''\in S_{N-1}} (-1)^{Q''} \sum_{L_1+V= \underline{N-1}}
(-1)^{|L_1|+|L|}  \prod_{v \in V} B_{n_v, 0_{Q''(v)}}(x, t)
\end{split}
\label{eq:g1l_onlyB}
\end{equation}
where we have rewritten $L=L_1+L_2$ with $L_1 \cap L_2 = \varnothing$  and  introduced the subset $V= L_2 + R$, which therefore allows us to write together the two products of $B$ integrals. Finally, we want to switch the two sums in order to obtain a Leibniz representation of a determinant using the relation~\cite{art:deuretzbacher2016_num}
\begin{equation}
\sum_{Q'' \in S_{N-1}} \sum_{L_1+V = \underline{N-1}} \cdots = \sum_{L_1+V = \underline{N-1}} \sum_{P \in S_{V}}      \binom{|V|}{|R|} \cdots,
\label{eq:sumswap}
\end{equation}
where, since we swapped the sums, the constraint on the elements of $L_1$ has to be taken into account in the sum of the permutations which are only the ones that do not include the indices $l \in \{Q(2), \dots Q(j)\}$ such that $Q''(l) = l$, namely they belong to $V$. The binomial coefficient in Eq.~(\ref{eq:sumswap}) counts the number of different decompositions of $V$ in terms of $L_2$ and $R$. Therefore, Eq.~\eqref{eq:g1l_onlyB} becomes
\begin{equation}
\begin{split}
g_{<}^{(1, j)} (\mathbf{n} | x, t)  &=  \sum_{i=1}^N (-1)^{i+1}\phi_{0_i} (x, t) 
\sum_{L_1+V = \underline{N-1}}
(-1)^{|L|+|L_1|} \binom{|V|}{|R|} \sum_{P \in S_{V}} (-1)^{P}\prod_{v \in V} B_{n_v, 0_{P(v)}}(x, t), 
\end{split}
\label{eq:g1l_onlyB_fin}
\end{equation}
where we can identify the Leibniz representation of a determinant, but we have to properly define the matrix.  For this purpose, we replace the sums over subsets with sums over certain indices. 

The sum over $L_1+V = \underline{N-1}$ means that we need to sum over all the possible indices, which are in total $N-1$. However, to keep into account the indices that belong to $L_1$, we introduce the matrix
\begin{equation}
    C_{n_v, 0_{P(v)}} (x, t) =
 \begin{cases}
     \delta_{n_v, 0_{P(v)}}  &{\rm if}\,\, P(v) = v \\
    B_{n_v, 0_{P(v)}} (x, t) & \mathrm{otherwise}
    \end{cases}.
    \label{eq:matfin}
\end{equation}
To rewrite the sum over $L_1+V = \underline{N-1}$, we introduce the index  $d=|V|$, namely the cardinality of $V$, which also fixes the cardinality of $L_1$. Indeed,  we know that $|R| = N-j$ and $|L| = j-1$, then $L_1 =  \underline{N-1} - R - L_2 =  \underline{N-1} - V$ and, therefore,  $|L_1| = N-1 - d$.  The index $d$ can assume the values $N-j, \dots, N-1$, because the cardinality of $V$ is given by $|L_2|+ |R| = |L_2| + N-j+1$ with $0 \leq |L_2| \leq |L| = j-2$. Once we sum over $d$, we need to take into account all the possible ordered sets, $(p_1, \dots, p_d)$, with size $d$, which can contain all the indices $\{2, \dots, N\}$  and, consequently, we have to sum over the elements of these sets.
Therefore, Eq.~\eqref{eq:g1l_onlyB_fin} becomes
\begin{equation}
\begin{split}
g_{<}^{(1, j)} (\mathbf{n} | x, t)  &= (-1)^{j+1}\sum_{i=1}^N (-1)^{i+1}\phi_{0_i} (x, t)  \sum_{d=N-j}^{N-1} (-1)^{N-1-d} \binom{d}{N-j}  \sum_{p \in P(d, N-1-d) }  \mbox{det}[(C_{n_k, 0_l})_{k, l \neq i \in p}].  
\end{split}
\label{eq:g1l_onlyB_det}
\end{equation}
where $P(d, N-1-d)$ includes the ordered sets $(p_1, \dots, p_d)$ and  $(q_1, \dots, q_{N-1-d})$, which are part of $V$ and $L_1$, respectively. This means that the indices $n_k, 0_l$ of the  $C$ matrix  are chosen in both the sets and when $l \in (q_1, \dots, q_{N-1-d})$, the matrix element is a $\delta_{n_k, 0_l}$. For all the rest, the matrix element is a $B_{n_k, 0_l}$.

Finally, inserting Eq.~\eqref{eq:g1l_onlyB_det} in Eq.~(9)
of the main text, the single-component lesser Green function is given by
\begin{equation}
\begin{split}
\mp i G^{<}_\sigma (x, t , y, t') &= \sum_{\mathbf{n}} \sum_{i, j =1}^{N} (\mp 1)^{i+j} 
[\eta^{(1, j)}_< (\mathbf{n} | \sigma, t')]^\ast
\eta^{(1, i)}_< (\mathbf{n} | \sigma, t) 
\sum_{a, b=1}^N (-1)^{a+b} \phi_{0_b}^\ast (y, t) \phi_{0_a} (x, t) \\
&\sum_{d=N-i}^{N-1} \sum_{e=N-j}^{N-1} (-1)^{d+e} \binom{d}{N-i}  \binom{e}{N-j} \\&\sum_{p \in P(d, N-1-d)} \sum_{q \in P(e, N-1-e)}  \mbox{det}[(C_{0_v, n_u})_{u, v \neq a \in p}(y, t')] \mbox{det}[(C_{n_l, 0_m})_{l, m \neq b \in q}(x, t)],
\end{split}
\label{eq:glesser_final}
\end{equation}
where the minus (plus) sign of $(\mp 1)^{i+j}$ is for fermions (bosons). We notice that the size of the two matrices is different if $i \neq j$ (to remember, $d$ and $e$ are the sizes of two different index subsets: the first one made of $R^{(i)} = \{Q(i+1), \dots, Q(N)\}$ and $L_2^{(i)} \in L^{(i)} = \{ Q(1), \dots, Q(i)\}$, and the second of $R^{(j)} = \{Q(j+1), \dots, Q(N)\}$ and $L_2^{(j)} \in L^{(j)} = \{ Q(1), \dots, Q(j)\}$. Therefore, we cannot write the product of  the two determinants as a determinant of the product of two matrices except in the case $i=j$, as instead it is possible in Refs.~\cite{art:settino2021, art:PatuFoerster2024}.

\section{Greater Green's function: Spatial part of form factors}\label{app:spatial_greater}
In this section, we derive a compact form to calculate Eq.~\eqref{eq:gn_1j_bar} as done in Sec.~\ref{app:spatial_lesser}. We write the determinant in Leibniz representation and remove a line from the $\left(N+1\right) \times \left(N+1\right)$ matrix, such that 
\begin{equation}
\begin{split}
g_>^{(1, j)} (\mathbf{m} | x, t) &= \frac{1}{(j-1)! (N-j+1)!}\sum_{i=1}^{N+1} (-1)^{i+1}\phi_{m_i}^\ast (x, t)\\
&\int_{x_2, \cdots, x_{j} < x < x_{j+1}, \cdots, x_{N+1}}\!\!\!\!\!\!\!\!\!\!\!\! d\vec{X}_N \sum_{Q, Q' \in S_{N}} (-1)^{Q+Q'} \left(\prod_{l=2}^{N+1} \phi_{Q(m_l)} (x_l, t) \right) \left(\prod_{k=2}^{N+1} \phi_{Q'(0_{k\neq i})} (x_{k\neq i}, t) \right).
\end{split}
\label{eq:g1l_bar}
\end{equation}
where we have divided by the total possible permutations in the two intervals: $[2, j]$ and $[j+1, N+1]$. Using Eqs.~\eqref{eq:Aij_g} and~\eqref{eq:Bij_g}, we obtain
\begin{equation}
\begin{split}
g_>^{(1, j)} (\mathbf{m} | x, t) &= \frac{1}{(j-1)! (N-j+1)!} \sum_{i=1}^{N+1} (-1)^{i+1}\phi_{m_i}^\ast (x, t)\sum_{Q, Q' \in S_N} (-1)^{Q+Q'} A_{Q(m_2), Q'(0_2)}(x, t)
\cdots
A_{Q(m_{j}), Q'(0_{j})}(x, t) \\
&\times B_{Q(m_{j+1}), Q'(0_{j+1})}(x, t) \cdots B_{Q(m_{N+1}), Q'(0_{N+1})}(x, t),
\end{split}
\label{eq:g1l_AB_bar}
\end{equation}
with  $\{m_2, \dots, m_{N+1}\} \neq m_i$. In the next step we introduce the permutation $Q''$ such that $Q'=Q'' \circ Q$ and the subsets $L=L_1 + L_2$ (same as for the lesser Green function) with cardinality $|L| = j-1$ and $\bar{R} = \{Q(j+1), \dots, Q(N+1)\}$ which contains an additional element with respect to $R$. Therefore Eq.~(\ref{eq:g1l_AB_bar}) becomes
\begin{equation}
\begin{split}
g_>^{(1, j)} (\mathbf{m} | x, t) &=\sum_{i=1}^{N+1} (-1)^{i+1}\phi_{m_i}^\ast (x, t) \sum_{Q''} (-1)^{Q''} \sum_{L+\bar{R} = \underline{N}} \prod_{l \in L} A_{m_l, 0_{Q''(l)}}(x, t) \prod_{r \in \bar{R}} B_{m_r, 0_{Q''(r)}}(x, t),
\end{split}
\label{eq:g1l_AB_1_bar}
\end{equation}
where we have multiplied the result by $(j-1)! (N-j+1)!$ because the sum over $L+\bar{R}$ contains  $(j-1)! (N-j+1)!$ and the ordering among $A$s and $B$s is not important~\cite{art:deuretzbacher2016_num}.

Using Eq.~\eqref{eq:prodAB}, Eq.~\eqref{eq:g1l_AB_1_bar} becomes
\begin{equation}
\begin{split}
g_>^{(1, j)} (\mathbf{m} | x, t) &= \sum_{i=1}^{N+1} (-1)^{i+1}\phi_{m_i}^\ast (x, t) 
\sum_{L_1+\bar{V} = \underline{N}}
(-1)^{|L|+|L_1|}  \binom{|\bar{V}|}{|\bar{R}|} \sum_{P \in S_{\bar{V}}} (-1)^{P}
\prod_{v \in \bar{V}} B_{m_v, 0_{P(v)}}(x, t)
\end{split}
\label{eq:g1l_onlyB_bar}
\end{equation}
where  we have introduced the new subset $\bar{V}= L_2 + \bar{R}$ and swapped the two sums using Eq.~\eqref{eq:sumswap}. We finally introduce the index $d=|\bar{V}|$, which can assume the values $N-j+1, \dots, N$, because $d= |L_2|+ |\bar{R}| = |L_2| + N-j+1$ with $0 \leq |L_2| \leq |L| = j-1$ and  Eq.~\eqref{eq:g1l_onlyB_bar} becomes
\begin{equation}
\begin{split}
g_>^{(1, j)} (\mathbf{m} | x, t) &=  (-1)^j\sum_{i=1}^{N+1} (-1)^{i+1}\phi_{m_i}^\ast (x, t)  \sum_{d=N-j+1}^{N} (-1)^{N+1-d} \binom{d}{N-j+1} \sum_{p \in P(d, N-d) }   \mbox{det}[(C_{m_k, 0_q})_{k\neq i, q\in p}], 
\end{split}
\label{eq:g1l_onlyB_det_bar}
\end{equation}
where we have introduced the matrix $C$, as defined in Eq.~\eqref{eq:matfin}, and $P(d, N-d)$ includes the ordered sets $(p_1, \dots, p_d)$ and  $(q_1, \dots, q_{N-d})$, which are part of $\bar{V}$ and $L_1$, respectively. 

Finally, inserting Eq.~\eqref{eq:g1l_onlyB_det_bar} in  
Eq.~(10)
of the main text, the $\sigma$-component greater Green function is given by
\begin{equation}
\begin{split}
i G^{>}_\sigma (x, t , y, t') &= \sum_{\mathbf{m}} \sum_{i, j =1}^{N+1} (\mp 1)^{i+j} [\eta^{(1, i)}_> (\mathbf{m} | \sigma, t)]^\ast
\eta^{(1, j)}_> (\mathbf{m} | \sigma, t') 
\sum_{a, b=1}^{N+1} (-1)^{a+b} \phi_{m_a}^\ast (x, t)\phi_{m_b} (y, t') \\
&\sum_{d=N-i+1}^{N} \sum_{e=N-j+1}^{N} (-1)^{d+e} \binom{d}{N-i+1}  \binom{e}{N-j+1}  \\&\sum_{p \in P(d, N-d) } \sum_{q \in P(e, N-e)}  \mbox{det}[(C_{0_v, m_u})_{v, u \neq a \in p}] \mbox{det}[(C_{m_l, 0_s})_{s, l \neq b \in q}],
\end{split}
\label{eq:ggreater_final}
\end{equation}
where the minus (plus) sign of $(\mp 1)^{i+j}$ is for fermions (bosons).

\section{Relation with the one-body density matrix}\label{app:1BDM}
In this Section, we show that the expression that we derived for the lesser Green function is consistent with the one for the one-body density matrix of Ref.~\cite{art:deuretzbacher2016_num}. Indeed, by definition, the lesser Green function at equal times corresponds to the one-body density matrix~\cite{book:mahan}, namely,
\begin{equation}
\begin{split}
\mp i G_\sigma^< (x, t, y, t) &= \rho_\sigma^{(1)} (x, y, t) = \sum_{i, j = 1}^N (\pm 1)^{i+j} \rho^{(i, j)} (x, y, t) c_\sigma^{(i, j)}(t).
\end{split}
\label{eq:G<rho1}
\end{equation}
Using Eq.(7)
of the main text
at $t=t'$ and the fact that
\begin{equation}
\begin{split}
& \sum_{\mathbf{n}} \braket{\vec{Y}_{N-1}, \vec{\tau}_{N-1} | e^{-i\hat H t} | \mathbf{n}} \braket{ \mathbf{n} | e^{i\hat H t} |\vec{X}_{N-1}, \vec{\sigma}_{N-1}} = \delta_{\vec{\sigma}_{N-1}, \vec{\tau}_{N-1}} \delta(\vec{X}_{N-1}-\vec{Y}_{N-1}), 
\end{split}
\label{eq:complete}
\end{equation}
due to the completeness relation of the energy eigenstates $\ket{\mathbf{n}}$, 
the spin part of the one-body density matrix, using the notation of Ref.~\cite{art:musolino2024}, is given by
\begin{equation}
\begin{split}
c_\sigma^{(i, j)} (t) &= \sum_{\mathbf{n}_\sigma} [\eta^{(1, i)}_< (\mathbf{n}_\sigma | \sigma, t)]^\ast
\eta^{(1, j)}_< (\mathbf{n}_\sigma | \sigma, t)\\
&=\sum_{\mathbf{n}_\sigma} \sum_{\vec{\sigma}_{N-1}} \sum_{\vec{\tau}_{N-1}}
\braket{\chi^{(0)}(t) | P^{-1}_{(1, \dots, i)} | \sigma, \vec{\tau}_{N-1}}\braket{\vec{\tau}_{N-1} | \chi^{(n)}(t)}\braket{\chi^{(n)}(t) | \vec{\sigma}_{N-1}}   \braket{\sigma, \vec{\sigma}_{N-1}| \hat P_{(1, \dots, j)} | \chi^{(0)}(t)}\\
&=\sum_{\vec{\sigma}_{N-1}} \braket{\chi^{(0)}(t) | \sigma, \vec{\sigma}_{N-1}}  \braket{\sigma, \vec{\sigma}_{N-1} | \hat P_{(i, \dots, j)} | \chi^{(0)}(t)}
\end{split}
\label{eq:cij_etas}
\end{equation}
and the spatial part for $x < x'$
\begin{equation}
\begin{split}
\rho^{(i, j)} (x, y, t) &=  \sum_{\mathbf{n}_X}
[g_{<}^{(1, i)} (\mathbf{n}_X | x, t)]^\ast
g_{<}^{(1, j)} (\mathbf{n}_X | x, t)\\
&= N! \int_{x_2 < \cdots x_{i-1} < x < x_{i+1} < \cdots <  x_{j-1} < y < x_{j} < \cdots < x_N} d\vec{X}_{N-1} [\Psi_F^{(0)} (y, \vec{X}_{N-1}, t)]^\ast
\Psi_F^{(0)} (x, \vec{X}_{N-1}, t),
\end{split}
\label{eq:rhoij_gn1j}
\end{equation}
where we have factorized the indices $\mathbf{n} = \mathbf{n}_X \otimes \mathbf{n}_\sigma$ without loss of generality. 

\section{Spin bandwidth}
Here we detail the procedure we have used in order to estimate the spin bandwidths $\Delta\omega^{({\bf n})}_{\rm B,F}$ shown in Fig.~3 of the main text. 
For each band, we have estimated
the maximum and the minimum values of the excitation energy 
to order $1/g$, by calculating the relevant spin-chain energies
as follows. 
For bosons, as already pointed out in the main text, there is no spin dispersion in the lower-energy  branches $\hbar \omega < \mu$, but 
it is observed in the upper-energy branches $\hbar \omega > \mu$. There, in each branch, the spectral peaks can be associated to either one of the following 
two spin symmetries:
 the fully-symmetric configuration ($N+1$) and the 
slightly less symmetric
one, ($N,1$).
The highest spin-chain energy 
is brought from the most symmetric diagram and its value is
of the order of
$\epsilon_\mathrm{B}^\mathrm{max}=2N {J}^{({\bf n})}_{c}/\hbar$, corresponding to $N$ symmetric exchanges of $N+1$ identical bosons. The 
smallest spin-chain energy
is brought by the highest-energy state
in the $N$-degenerate manifold of the symmetry sector ($N,1)$.
An upper bound can be obtained by assuming
that in such a state the energy 
is as if $N-2$ identical bosons contribute (i.e. the vertical column with two boxes does not contribute), leading to $\epsilon_\mathrm{B}^\mathrm{min}=2(N-2) {J}^{({\bf n})}_{c}/\hbar$.  We have verified this assumption through the exact diagonalization of the spin-chain  Hamiltonian.
The above estimate allows us to obtain for the bosonic upper-energy spin bandwidths $\Delta\omega^{({\bf n})}_{\rm B}= \epsilon_\mathrm{B}^\mathrm{max} - \epsilon_\mathrm{B}^\mathrm{min} \simeq 4{J}^{({\bf n})}_{c}/\hbar$.

The evaluation of
 bandwidth of the fermionic mixture is less straightforward,
because of the 
mixed symmetry of the Young diagram $(2,2,\dots,1)$ involved both in the lower ($\hbar \omega < \mu$) and upper ($\hbar \omega > \mu$) branches of the spectrum, the only difference being
the number of two-boxes lines between the two cases: $N/2-1$ for the lower-energy branches and $N/2$ for the upper-energy ones.
We start from the observation that in a homogeneous system with periodic boundary conditions, the energy of the spin state
depends on an orbital quantum number that determines the total momentum of the mixture on the ring \cite{AupetitDiallo2024}.
For such mixed-symmetry
diagrams, the orbital quantum number selects particular spin-state configurations. The highest spin-chain energy
comes from the 
N\'eel configuration $|\uparrow\downarrow\uparrow\downarrow\dots\uparrow\rangle$
and its value is $\epsilon_\mathrm{F-}^\mathrm{max}=(N-2){J}^{({\bf n})}_{c}/\hbar$ for the lower-energy branches and $\epsilon_\mathrm{F+}^\mathrm{max}=N {J}^{({\bf n})}_{c}/\hbar$ for the upper-energy ones
corresponding to $N-1$ ($N+1$) symmetric exchanges of  distinguishable spins.
The lowest spin-chain  energy  can be estimated using
a domain-wall configuration $|\uparrow\uparrow\dots\uparrow\downarrow\dots\downarrow\downarrow\rangle$ and thus is of the order \cite{footnote}
 The different scaling of $\epsilon_\mathrm{F\mp}^\mathrm{max}$  and $\epsilon_\mathrm{F}^\mathrm{min} $ with particle number 
 is the crucial point at the basis of the fact that fermionic mixtures have a  larger spin bandwidth than 
bosonic ones.
Indeed, for the bandwidth of the lower/upper energy spin branches of the fermionic spectrum  we find that
$\Delta\omega^{({\bf n})}_{\rm F}= \epsilon_\mathrm{F\mp}^\mathrm{max} -\epsilon_\mathrm{F}^\mathrm{min}\simeq (N-1\mp2) {J}^{({\bf n})}_{c}/\hbar$.
We finally notice that the above calculations are approximate estimates since for systems in an external trap neither N\'eel nor domain-wall spin configurations are eigenstates of the spin chain Hamiltonian. 
We have however verified that, even in the presence of the harmonic confinement, these two spin configurations 
bring the main contribution to the eigenstates corresponding to the highest and smallest spin excitation energy, respectively.

\end{widetext}

\end{document}